\documentclass{PoS}
\usepackage{graphicx}
\PoS{PoS(LAT2005)183}

\title{QCD Level Density from Maximum Entropy Method }
\ShortTitle{QCD level density from maximum entropy method }

\author{\speaker{Shinji Ejiri} \\
        Department of Physics, University of Tokyo, Tokyo 113-0033, Japan\\
        E-mail: \email{ejiri@phys.s.u-tokyo.ac.jp}}

\author{Tetsuo Hatsuda\thanks{Supported by Grants-in-Aid of the Japanese 
 Ministry of Education, Culture, Sports, Science and Technology, No. 15540254.} \\
        Department of Physics, University of Tokyo, Tokyo 113-0033, Japan\\
        E-mail: \email{hatsuda@phys.s.u-tokyo.ac.jp}}

\abstract{
We propose a method to calculate the QCD level density directly from 
the thermodynamic quantities obtained by lattice QCD simulations with 
the use of the maximum entropy method (MEM).  
Understanding QCD thermodynamics from QCD
spectral properties  has its own importance. Also it has
a close connection to phenomenological analyses of
the lattice data as well as experimental data
on the basis of hadronic resonances. Our feasibility study shows that the 
MEM can provide a useful tool to study QCD level density. }

\FullConference{XXIIIrd International Symposium on Lattice Field Theory\\
		25-30 July 2005\\
		Trinity College, Dublin, Ireland}

\begin{document}

\section{QCD level density}
\label{sec:intro}

In this report, we discuss the QCD level density 
in a finite box with a size $V$. Denoting energy eigenstates 
of the QCD Hamiltonian ${\cal H}_{QCD}$ by $E_n$, 
the energy-level density $A(E,V)$ is defined by counting the number of 
states in the range of $E$ to $E +dE$, i.e., 
\begin{eqnarray}
A(E,V)  \equiv \sum_n \delta(E-E_n) \ \ \ {\rm with} \ \ \ 
{\cal H}_{QCD} |\Psi \rangle = E_n |\Psi \rangle . 
\end{eqnarray}
Using this level density, the partition function is written as
 its Laplace transform; 
\begin{eqnarray}
{\cal Z}(\beta,V) = {\rm Tr} \left[ e^{-\beta {\cal H}_{QCD}} \right] 
= \int_0^{\infty} A(E,V) e^{-\beta E} dE,
\label{eq:partition} 
\end{eqnarray}
where $\beta \equiv 1/T$ with $T$ being the temperature of the system.
Thermodynamic quantities such as the pressure, $p$, and the 
energy density, $\varepsilon$, are related to 
the partition function as
\begin{eqnarray}
p(\beta) = \frac{1}{V \beta} \ln {\cal Z} (\beta, V), \hspace{1cm} 
\varepsilon(\beta) = -\frac{1}{V} 
\frac{\partial \ln {\cal Z} (\beta, V)}{\partial \beta}.
\label{eq:prs} 
\end{eqnarray}
Eq.(\ref{eq:partition}) together with Eq.(\ref{eq:prs})
imply that every thermodynamic quantity is obtained from 
the information of the level density defined at zero temperature.
Especially, Eq.(\ref{eq:partition}) tells us that
the properties of QCD phase transition at finite $T$,
which is characterized by the non-analytic behavior of $p(\beta)$
in the thermodynamic limit, are already encoded in the QCD level
density defined at $T=0$.

The investigation of the level density 
in nuclear physics and in hadron physics has a long history:
For example, 
Bethe studied the nuclear level density by evaluating
the partition function of the Fermi gas \cite{bethe}.
By performing the inverse Laplace transform, he showed
that $A \propto \exp(2 \sqrt{c E})$ with $c$ being a
constant related to single-particle level density.
Hagedorn studied hadronic level density and has derived 
an asymptotic formula for the state  density of hadrons 
with a  mass $m$, 
\begin{eqnarray}
\rho (m) \propto \frac{1}{m^{a}} \exp (m/T_H) . 
\label{eq:hag} 
\end{eqnarray} 
Here $a$ and $T_H$ are some constants and the latter is 
called the Hagedorn temperature \cite{hage}.
Eq.(\ref{eq:hag}) is derived from the celebrated bootstrap model in which 
the state density $\rho (m)$ and the energy-level density $A(E=m, V_0)$ 
in a fireball of size $V_0$  are identified for large $m$. 
The exponential growth of the hadronic state density, 
Eq.(\ref{eq:hag}), agrees with experimental data up to 2 GeV 
at present and the agreement becomes better as new resonances are 
included \cite{boroni04}. Because of this success, 
the hadron resonance gas model as well as the Hagedorn's formula
have been and is being widely used in QCD phenomenology. 
However, this model describes only the hadronic matter at 
temperature below $T_H$. 
If we try to calculate ${\cal Z}$ with Eq.(\ref{eq:hag}) above $T_H$, 
the integral does not converge. 

On the other hand, progresses in QCD thermodynamics 
has been obtained steadily by the first principle lattice simulations.
The results show that $\varepsilon$ and $p$ 
increase rapidly near the transition temperature $T_c$ and 
approach to the black-body formula, $\varepsilon \sim 3 p \propto T^4$,
for sufficiently high temperature. 
It has been also found that thermodynamics below $T_c$ is well described 
by hadron resonance gas model as long as appropriate hadron masses 
relevant to lattice simulations are employed \cite{KRT,BS05}.  
Therefore, the time is now ripe to consider a unified and model-independent 
description of QCD thermodynamics on the basis of the QCD level density.
In fact, detailed studies of the level density
is particularly useful to identify the relevant degrees of freedom 
in hot QCD below and above $T_c$.

Now let us first pay attention to a close similarity
between Eq.(\ref{eq:partition}) and the spectral representation of the 
hadronic two-point correlation function, $G$, as a function of the 
imaginary time $\tau$:
\begin{eqnarray}
G(\tau) 
= \int_0^{\infty} R(\omega) e^{-\tau \omega} d\omega, 
\label{eq:spf} 
\end{eqnarray}  
where $R(\omega)$ is the spectral function and $\omega$ is the frequency.
Formal correspondences, ${\cal Z} \leftrightarrow G$,
$\beta \leftrightarrow \tau$,
$E \leftrightarrow \omega$ and $A \leftrightarrow R$,
are clear by comparing Eq.(\ref{eq:partition}) with Eq.(\ref{eq:spf}).
Since the maximum entropy method (MEM) is known to be a powerful 
tool to extract the spectral function $R$ from the lattice data 
$G$ \cite{MEM}, the same technique is expected to be used to extract 
the level density $A(E,V)$ from the lattice data of ${\cal Z}$.

\section{A toy model}
\label{sec:model}

\begin{figure}[t]
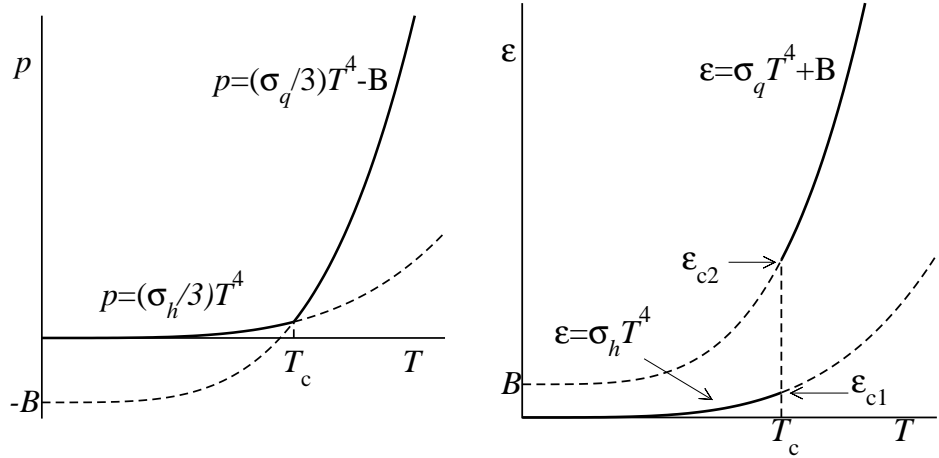

\begin{center}
\includegraphics[width=2.3in]{pvst_bagm.eps}
\hskip 0.6cm
\includegraphics[width=2.3in]{evst_bagm.eps}
\caption{A schematic model for the pressure (left panel) and the energy 
density (right panel).}
\end{center}
\setcounter{figure}{1}
\end{figure} 

Before testing the idea of using MEM to extract the level density $A(E,V)$,
let us first discuss the general structure of $A$ expected in
simple cases where some analytic study is possible.
First of all, the inverse Laplace transform (the Bromwich integral)
of Eq.(\ref{eq:partition}) reads 
\begin{eqnarray}
A(E,V) = \frac{1}{2\pi i} \int_{\gamma -i \infty}^{\gamma +i \infty}  
{\cal Z} (\beta, V) e^{E \beta} d\beta 
= \frac{1}{2\pi i} \int_{\gamma -i \infty}^{\gamma +i \infty}  
  e^{(p(\beta) V + E) \beta } d\beta , 
\label{eq:brom} 
\end{eqnarray}
where $\gamma$ is a real number chosen so that all the 
singularities of ${\cal Z} (\beta, V)$ are to the left of it.
In the leading order of the saddle point approximation, 
one readily finds
\begin{eqnarray}
A(E,V) \propto \exp (s (\varepsilon ) V),
\label{eq:saddle}
\end{eqnarray}
where $\varepsilon (\equiv E/V)$ is the energy
density and $s(\varepsilon)$ is the entropy density.  

Consider a simplest case of free massless particles
where the pressure is given by $p(\beta)=(\sigma/3) T^4$.
Then, by working out the Gaussian integral around the 
saddle point, one finds
\begin{eqnarray}
\left. A(E,V) \right|_{EV^{1/3} \gg 1}
   \sim \frac{1}{\sqrt{8 \pi V}}
   \left( \frac{\sigma}{(E/V)^5} \right)^{1/8} 
     \exp \left( \frac{4}{3} \sigma^{1/4} V (E/V)^{3/4} \right) ,
\label{eq:free}
\end{eqnarray}
which implies  $\ln A \sim E^{3/4}$ at high excitation energies.

To examine a more realistic example with a phase transition, 
let us consider the bag equation of state in which  
we assume free massless pions (free massless quarks and gluons)
below (above) $T_c$.
In this case, the pressure of the system is given by 
\begin{eqnarray}
p(T) = \frac{T}{V} \ln {\cal Z}
= \frac{\sigma_h}{3} T^4 \theta (T_c -T) 
+ \left( \frac{\sigma_q}{3} T^4 -B \right) \theta (T -T_c), 
\label{eq:pbag} 
\end{eqnarray}
where $B$ is the bag constant which is
related the critical temperature as $T_c = [(3B)/(\sigma_q - \sigma_h)]^{1/4}$. 
$\sigma_h$ and $\sigma_q$ are proportional to the number of degrees of
freedom in the hadronic phase and the quark-gluon plasma phase, respectively. 
A schematic sketch for $p/T^4$ and $\varepsilon /T^4$ is given in Fig.~1.

For the bag equation of state given in Eq.(\ref{eq:pbag}),
the exponent $s(\varepsilon)$ in Eq.(\ref{eq:saddle}) turns out to be
\begin{eqnarray}
\frac{4}{3} \sigma_h^{1/4} \varepsilon^{3/4} 
\hspace{3mm} {\rm [I]}, \hspace{7mm} 
 \frac{1}{3} \sigma_h T_c^3 + \frac{\varepsilon}{T_c} 
\hspace{3mm} {\rm [II]}, \hspace{7mm} 
 \frac{4}{3} \sigma_q^{1/4} \left( \varepsilon -B \right)^{3/4} 
\hspace{3mm} {\rm [III]}.
\label{eq:lna} 
\end{eqnarray}
Here the formulas [I], [II] and [III] are valid for 
$\varepsilon < \varepsilon_{c1}$, 
$\varepsilon_{c1} < \varepsilon < \varepsilon_{c2}$, and 
$\varepsilon > \varepsilon_{c2}$, respectively. 
Note that  $\varepsilon_{c1} = \sigma_h T_c^4$ 
($\varepsilon_{c2} = \sigma_q T_c^4+B 
= \frac{1}{3} \left(4 \sigma_q - \sigma_h \right) T_c^4)$
is the energy density just below (above) the phase transition point
as shown in the right panel of Fig.~1.

The behavior of the exponent $s(\varepsilon)$  given in Eq.(\ref{eq:lna}) 
is illustrated in the left panel of Fig.~2. 
The level density shows the Hagedorn type behavior $\ln A \propto E$
in the phase transition region [II], while
it shows softer behavior $\ln A \sim E^{3/4}$ 
in the high temperature region [III].
The exponent has a crossover from $E$ to $E^{3/4}$ 
at $E=E_{c2}=\varepsilon_{c2} V$ at which the
quark-gluon plasma starts to emerge.

\begin{figure}[t]
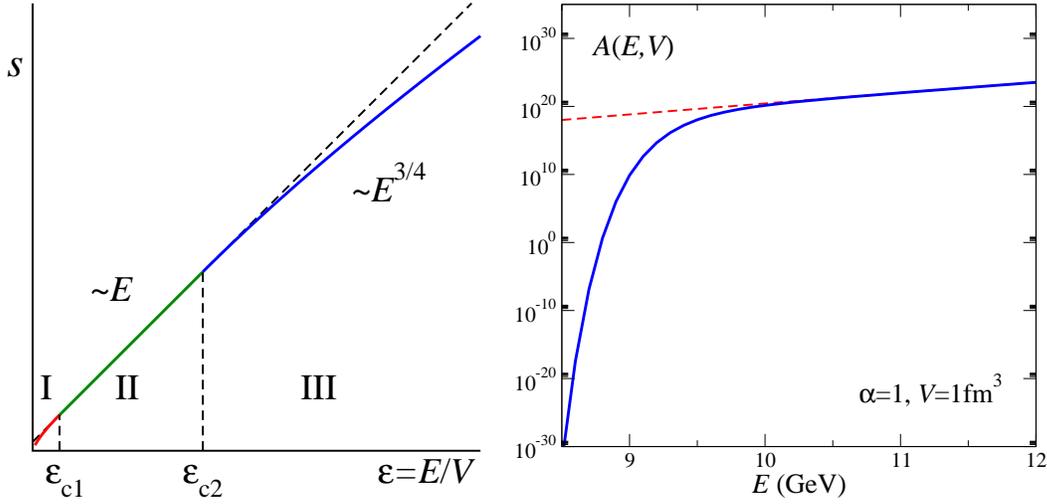

\begin{center}
\includegraphics[width=2.5in]{lna_fig.eps}
\hskip 0.3cm
\includegraphics[width=2.8in]{AvsE_comAs.eps}
\caption{Left panel: A schematic picture of 
$s(\varepsilon) \propto \ln A (E,V)$ in a toy model
as a function of $\varepsilon =E/V$. 
 Right panel: The QCD level density $A(E,V)$ 
in an arbitrary unit extracted from the lattice data.}
\end{center}
\setcounter{figure}{2}
\end{figure}

\section{Application of MEM}
\label{sec:method}

Let us now try to extract $A(E,V)$ from
lattice QCD data of the thermodynamic quantities. 
Following the idea of MEM,
we minimize the ``free-energy'' functional $Q(A)$ with respect to the
level density $A$: 
\begin{eqnarray}
Q = \frac{\chi^2}{2} - \alpha S_{\rm inf}.
\label{eq:qmem} 
\end{eqnarray}
Here, $S_{\rm inf}$ is the information entropy given by 
\begin{eqnarray}
S_{\rm inf} = \int (f-1 -f \ln f) dE, \hspace{5mm} 
A(E, V) = A_{\rm asy}(E,V) \times f(E,V),
\label{eq:sjentropy} 
\end{eqnarray}
where $A_{\rm asy}(E,V)$ implies a  ``default model''
representing the asymptotic behavior of $A$ in the large $E$ limit. 
In the present study, we adopt
$A_{\rm asy}(E,V) = ({V^{1/3}}/{\sqrt{8 \pi}}) 
\exp \left( \frac{4}{3} \sigma^{1/4} V\epsilon^{3/4} \right)$
which is proportional to the asymptotic form of $A$ for
a free quark-gluon gas. We adjust 
$\sigma$ to reproduce $\varepsilon / T^4$ measured on the 
lattice at high temperature. 
For the likelihood function $\chi^2$, we have chosen
\begin{eqnarray}
\chi^2 = \sum_{i=1}^{n} \left( 
\frac{ p_i^{\rm lat} - \frac{1}{V \beta_i} \ln \int A(E,V) e^{-\beta_i E} dE}
{\Delta p_i^{\rm lat}} \right)^2
+ \sum_{i=1}^{n} \left(
\frac{ \varepsilon_i^{\rm lat} 
-\frac{\int (E/V) A(E, V) e^{-\beta_i E} dE}{\exp(V \beta_i p_i^{\rm lat}) }}
{\Delta \varepsilon_i^{\rm lat}} \right)^2,
\label{eq:likelihood} 
\end{eqnarray}
where $p_i^{\rm lat}$ and $\varepsilon_i^{\rm lat}$ are the data obtained by 
lattice simulations at $n$ discrete values of the inverse temperature
$\beta_i$.
Also, $\Delta p_i^{\rm lat}$ ($\Delta \varepsilon_i^{\rm lat}$) denotes
statistical error of the pressure (energy density).

We use the 2-flavor full QCD data with improved Wilson quarks generated on 
a $16^3 \times 4$ lattice \cite{cppacs01}. The 
spatial size of the lattice in the physical unit is approximately
$(4 {\rm fm})^3$, although the physical volume changes for different 
values of  $\beta_i$. 
As a first step, we choose $V=1 $ fm$^3$ in Eq.(\ref{eq:likelihood})
by assuming small volume-dependence of the lattice data. 
We use the data for the pressure and energy density 
with $m_{PS}/m_{V}=0.90$ in which there are 
``seven'' independent data points ($n=7$).

$A(E,V)$ reconstructed by MEM is shown by the solid line in the 
right panel of Fig.2 in an arbitrary unit with the logarithmic scale. 
Dashed line is $A_{\rm asy}$ mentioned above.
To set the  scale in the horizontal axis, we assume 
$T_c \approx 175 {\rm MeV}$. Also,  
we have chosen $\alpha =1$ in the present MEM analysis: eventually
it has to be eliminated by calculating the probability distribution
$P[\alpha]$ \cite{MEM}.
As can be seen from the figure, the behavior of $A(E,V)$ at high $E$ is
consistent with $A_{\rm asy} \sim E^{3/4}$ (the dashed line), while
$A(E,V)$ decreases strongly around $E=9-10 {\rm GeV}$ and 
deviates substantially  from  $A_{\rm asy}$ at low energies. 
This rapid crossover of $\ln A$ is qualitatively consistent with
what we have discussed using the toy model.

Now we briefly discuss some systematic uncertainty in the MEM analysis.  
Instead of utilizing both $p_i^{\rm lat}$ and $\varepsilon_i^{\rm lat}$
as in Eq.(\ref{eq:likelihood}), 
we have done MEM analyses by using  $p_i^{\rm lat}$ only 
and by using  $\varepsilon_i^{\rm lat}$ only.
In these cases, the crossover  
region has moved $\pm 20\%$ from that shown in the right panel of  Fig.~2.
Therefore, the systematic error due to different choices of $\chi^2$
is still large at present as long as we take only seven data points.
Nevertheless, we believe that MEM could become a useful tool to extract 
the QCD level density if large number of data points with high accuracy 
become available in the future.

\section{Conclusions and outlook}
\label{sec:conclusion}

To understand thermodynamic properties near the QCD phase 
transition, the knowledge of the energy-level density 
is quite useful. In this report,
we proposed a new method to calculate the QCD level density 
and have made a feasibility test using lattice data.
Although further systematic studies are necessary by increasing the number of
data points, the direct calculation of the QCD level density seems 
to be possible using the maximum entropy method. 

An extension of the present study to the system at non-zero baryon density 
is interesting to be explores in connection with  recent progress of 
finite density lattice QCD. 
The grand partition function ${\cal Z}(\beta,\mu,V)$ can be separated 
into the canonical partition functions ${\cal Z}_N(\beta,V)$ for 
each fixed quark number $N$ by the fugacity expansion as 
\begin{eqnarray}
{\cal Z}(\beta, \mu, V) 
= \sum_N e^{N \beta \mu} {\cal Z}_N(\beta,V) 
= \sum_N e^{N \beta \mu} \int_0^{\infty}  A(E,N,V) e^{-\beta E} dE. 
\label{eq:nonzeromu} 
\end{eqnarray}
Then, by using the lattice data with several different values of 
$\beta$ and $\mu$, one may extract the level density $A(E,N,V)$. 
Such an analysis will shed lights on  
colored composite states with non-zero baryon numbers above $T_c$ 
such as  the quark-gluon bound state and the diquark \cite{shur,EKR}. 
If such states are important to thermodynamic quantities,
they should also show up in the  QCD level densities with $N=1$ and $N=2$.


\begin{thebibliography}{99}


\bibitem{bethe} H.A.~Bethe, \emph{An Attempt to Calculate the 
Number of Energy Levels of a Heavy Nucleus},
\emph{Phy.~Rev.} {\bf 50} (1936) 332. 
For further references, see e.g., H. Ahmodov, I. Zorba, M. Yilmaz and
 B. G\"{o}n\"{u}l, \emph{On the level density of even-even
  nuclei in the regions of rare-earth and actinide elements},
 \emph{Nucl. Phys. A} {\bf 706} (2002) 313.

\bibitem{hage} R.~Hagedorn, 
\emph{How we got to QCD matter from the hadron side by trial and error}, 
\emph{Lecture Notes in Physics} {\bf 221} (1985) 53.

\bibitem{boroni04}
W.~Broniowski, W.~Florkowski and L.Y.~Glozman, 
\emph{Update of the Hagedorn mass spectrum}, 
\emph{Phys.~Rev.~D}{\bf 70} (2004) 117503 
[{\tt hep-ph/0407290}]. 

\bibitem{KRT}
 F.~Karsch, K.~Redlich and A.~Tawfik,
\emph{Hadron resonance mass spectrum and lattice QCD thermodynamics}, 
\emph{Eur.~Phys.~J.~C}{\bf 29} (2003) 549 
[{\tt hep-ph/0303108}]; 
\emph{Thermodynamics at non-zero baryon number density: 
A comparison of lattice and hadron resonance gas model calculations}, 
\emph{Phys.~Lett.~B}{\bf 571} (2003) 67 
[{\tt hep-ph/0306208}].

\bibitem{BS05}
C.R.~Allton, M.~D\"{o}ring, S.~Ejiri, S.J.~Hands, O.~Kaczmarek, F.~Karsch, 
E.~Laermann and K.~Redlich, 
\emph{Thermodynamics of two flavor QCD to sixth order in quark 
chemical potential}, 
\emph{Phys.~Rev.~D}{\bf 71} (2005) 054508 
[{\tt hep-lat/0501030}].

\bibitem{MEM} 
Y.~Nakahara, M.~Asakawa and  T.~Hatsuda, 
\emph{Hadronic Spectral Functions in Lattice QCD}, 
\emph{Phy. Rev. D}{\bf 60} (1999) 091503 
[{\tt hep-lat/9905034}].\\
M.~Asakawa, T.~Hatsuda and Y.~Nakahara, 
\emph{Maximum Entropy Analysis of the Spectral Functions in Lattice QCD}, 
\emph{Prog.~Part.~Nucl.~Phys.}{\bf 46} (2001) 459 
[{\tt hep-lat/0011040}].

\bibitem{cppacs01}
A.~Ali Khan {\it et al.} (CP-PACS Collaboration), 
\emph{Equation of state in finite-temperature QCD with 
two flavors of improved Wilson quarks}, 
\emph{Phys.~Rev.~D}{\bf 64} (2001) 074510 
[{\tt hep-lat/0103028}].

\bibitem{shur}
E.V.~Shuryak and I.~Zahed, 
\emph{Towards a Theory of Binary Bound States in the Quark-Gluon Plasma}, 
\emph{Phys.~Rev.~D}{\bf 70} (2004) 054507 
[{\tt hep-ph/0403127}].

\bibitem{EKR}
S.~Ejiri, F.~Karsch and K.~Redlich, 
\emph{Hadronic fluctuations at the QCD phase transition}, 
[{\tt hep-ph/0509051}].

\end{thebibliography}
\end{document}